\documentclass[12pt]{aastex62}

\begin{document}
\raggedright
\huge
High Redshift Obscured Quasars and the Need for Optical to NIR, Massively Multiplexed, Spectroscopic Facilities\linebreak
 \linebreak
\normalsize

\noindent \textbf{Thematic Areas:} \hspace*{60pt} $\square$ Planetary Systems \hspace*{10pt} $\square$ Star and Planet Formation \hspace*{20pt}\linebreak
$\square$ Formation and Evolution of Compact Objects \hspace*{31pt} $\square$ Cosmology and Fundamental Physics \linebreak
  $\square$  Stars and Stellar Evolution \hspace*{1pt} $\square$ Resolved Stellar Populations and their Environments \hspace*{40pt} \linebreak
  $\boxtimes$    Galaxy Evolution   \hspace*{45pt} $\square$             Multi-Messenger Astronomy and Astrophysics \hspace*{65pt} \linebreak
  
\textbf{Principal Author:}

Name:	Andreea Petric
 \linebreak						
Institution:  Inst. for Astronomy, Canada France Hawaii Tel., Maunakea Spectroscopic Explorer
 \linebreak
Emails: apetric@hawaii.edu, petric@cfht.hawaii.edu
 \linebreak

\textbf{Co-authors:} \newline
{\bf{Mark Lacy}}, National Radio Astronomy Observatory;
{\bf{St\'ephanie Juneau}}, National Optical Astronomy Observatory;
{\bf{Yue Shen}}, University of Illinois at Urbana-Champaign;
{\bf{Xiaohui Fan}}, Steward Observatory, University of Arizona;
{\bf{Nicolas Flagey}}, Canada France Hawaii Telescope;
{\bf{Yjan Gordon}}, University of Manitoba;
 {\bf{Daryl Haggard}},  McGill University, McGill Space, CIFAR Azrieli Global Scholar, Gravity \& the Extreme Universe Program, Canadian Institute for Advanced Research;
{\bf{Patrick B. Hall}}, York University, Toronto, Canada;
{\bf{Nimish Hathi}}, Space Telescope Science Institute;
{\bf{Dragana Ilic}}, Department of astronomy, Faculty of mathematics, University of Belgrade;
{\bf{Claudia D.P. Lagos}}, International Centre for Radio Astronomy Research, U. of Western Australia;
{\bf{Xin Liu}}, University of Illinois at Urbana-Champaign;
{\bf{Christopher O'Dea}}, University of Manitoba;
{\bf{Luka {\v C}. {Popovi{\'c}}}}, Astronomical Observatory Belgrade;
{\bf{Andy Sheinis}}, Canada France Hawaii Telescope;
{\bf{Yiping Wang}}, National Astronomical Observatories, Chinese Academy of Sciences;
{\bf{Yongquan Xue}}, University of Science and Technology of China
\linebreak

\textbf{Abstract:}
Most bulge-dominated galaxies host black holes with masses that tightly correlate with the masses of their bulges. This may indicate that the black holes may regulate galaxy growth, or vice versa, or that they may grow in lock-step. 
The quest to understand how, when, and where those black-holes formed motivates much of extragalactic astronomy. Here we focus on a population of galaxies with active black holes in their nuclei (active galactic nuclei or AGN), that are fully or partially hidden by dust and gas: the emission from the broad line region is either completely or partially obscured with a visual extinction of 1 or above. This limit, though not yet precise, appears to be the point at which the populations of AGN may evolve differently. We highlight the importance of finding and studying those dusty AGN at redshifts between 1 and 3, the epoch when the universe may have gone through its most dramatic changes. We emphasize the need for future large multiplexed spectroscopic instruments that can perform dedicated surveys in the optical and NIR to pin down the demographics of such objects, and study their reddening properties, star-formation histories, and excitation conditions. These key studies will shed light on the role of black holes in galaxy evolution during the epoch of peak growth activity.
\pagebreak
\section{Introduction}

Observations of near and far massive galaxies with sufficient spatial and velocity resolution have revealed the dynamical signatures of black holes (BH) and the high luminosity signatures of growing massive BH through the accretion of gas and/or tidally disturbed stars.
Many questions still remain, however, and answering them is crucial to advancing our understanding of galaxy and supermassive black hole (SMBH) evolution.
To understand how, when, and where did all the present epoch super-massive black holes (SMBH) grow, astronomers have studied large samples of growing BH and computed accretion rates from estimated bolometric luminosities.
There are indications that a significant fraction of the growth of black holes took place during obscured phases \citep{kelly2010}.
An accurate census of accretion rates that includes dust obscured AGN is needed to compare with today's black hole density \citep{soltan1982, Lacy2015, hopkins2007, mart2009, del2014}.
\newline

Growing SMBHs power compact, luminous AGN, and their growth may be connected to the evolution of the host galaxy, or at least the host galaxy's bulge. It remains unclear if, and for which type of galaxies, this connection is accidental or causal, and the path to this correlation may be related to the nature of the sample studied, (e.g. radio loud AGN tend to be major mergers \citep{chia2015}), the bias of the detection technique \citep{jun2011}, or our lack of sufficiently large samples at redshifts when most of the growth activity happens).
We suspect co-evolution between AGN and their hosts because of the correlation between the central SMBH and the surrounding bulge stars \citep[e.g.][]{koh2013,jah2011,hopkins2008}. \citet{jah2011} have argued that galaxy mergers are able to produce galaxies with central SMBH of masses proportional to their galactic bulge masses. However, the scatter in this correlation is larger than measurement errors which suggests the need for other mechanisms at play: processes associated with the central accreting SMBH and/or massive star-formation (e.g. supernovae) that shape the evolution of the host galaxy (feedback). \newline
 
 
One proposed evolutionary path for the most luminous AGN, known as QSOs, involves a symbiotic relation between the growing SMBH and its host galaxy whereby they control each-other’s growth. Gas rich galaxies interact gravitationally, and as they merge their ISM evolves: gas gets compressed in the changing gravitational potential of the merging system increasing SFR; the gas then loses angular momentum, and a fraction of it falls toward the center and feeds the central SMBH. During this stage the broad line emission, from the aptly named broad line region — a sub-pc region around the SMBH, filled with hot gas, optically thick to ionizing continuum radiation — is absorbed and scattered by gas and dust. At this point in its evolution, such a system is lacking observable broad line emission and is observed as a type 2 QSO (QSO2). QSO2s show only narrow lines with widths a $\leq 10^3 $ km/s. The young stars and growing SMBH continue to consume the surrounding ISM and inject energy into it through shocks, radiation, and turbulence. Those processes increase the pressure in the gas, ionize part of it, and ultimately impede star-formation. As the obscuring material is cleared from the center, the broad-line emission (width of $\sim 10^4$ km/sec) can escape and be detected. Outflows impacting the interstellar-medium of the host galaxy have been studied in several nearby radio-loud AGN \citep[e.g][]{morganti2013,morganti2017}. Such a system would be observed as a type 1 QSO (QSO1). This evolutionary theory suggests that the observed properties of AGN are determined by the evolutionary stage at which we observe them. This “merger-to-QSO1” path naturally leads to positive correlations between the masses of galactic bulges and those of their central SMBHs: in the early stages, the SMBHs and the hosts grow together, and in the later stages the AGN consumes, pushes or puffs out the fuel, halting both accretion onto the SMBH and star-formation. \newline

An alternative explanation for the relationship between broad-line QSO1 and narrow-line QSO2s is the orientation theory \citep[e.g.][]{anton1993,mil91} which stipulates that the two classes include similar object and that the observed differences are due to our perspective as observers. This theory postulates that QSO1s and QSO2s contain a dusty torus, which absorbs the broad-line emission when viewed edge on (we thus see a QSO2) and allows the broad-line emission through when viewed face-on (we thus see a QSO1). \citet{elitzur2012} proposed a more sophisticated version of the unification theory of AGN: all AGNs may indeed have a torus, but this torus is clumpy, and both the viewing geometry and the number and properties of clumps along the line of sight dictate the observable characteristics of the majority of AGN. \citet{elitzur2012} describes a fundamental parameter space for AGN spanned by two independent axes: orientation and covering factor. Evidence for this comes from the observation of a source, which was initially identified as a QSO2 and, when the obscuring cloud in the torus moved away from the line of sight, was found to host a QSO1 \citep{Aretxaga1999}. Other support for this theory comes from spectro-polarimetric data showing that spectra of scattered light in QSO2s contain broad line emission (i.e. typical of QSO1, see e.g. \citet{zakamska2005}). Clumpy torus models were shown to fit well the X-ray to IR emission of lower luminosity type 1 and type 2 AGN consistent with the suggestions of \citet{elitzur2012}: see also \citet{ram2017}.
\newline

One way to potentially distinguish between these scenarios is by comparing the environments of QSO1s and QSO2s. The orientation theory would predict identical environments, whereas an evolutionary scenario may result in the average environment of a QSO2 differing from that of a QSO1. Observations to date have been ambiguous, with some studies finding a higher density of field galaxies around obscured AGN \citep{donoso2014}, and some finding results consistent with no difference \citep{koutoulidis2018} or higher densities around {\em unobscured} objects \citep{allevato2014}. These studies used AGN selected in different ways, and noisy estimates of the environment, however. Spectroscopic redshifts with a dense sampling rate are needed to obtain sufficiently precise estimates of the environment. 
\newline

\textbf{Spectroscopic studies of large samples of AGN, including the reddest and faintest, and their environments, are thus needed to disentangle the origins and evolution of nuclear obscuration in AGN.} 


\section{Dusty Quasars}
Recent studies discovered that the farthest, most luminous, and dustiest of quasars bear the marks of gravitational interactions from tidal tails to complex nuclear structure. \citet{lacy2018, glikman2015, urrutia2008} use HST and ground base adaptive optics to image obscured AGN, separate the emission from the nucleus and the host, and find that the hosts of obscured AGN tend to be major mergers. The observations of \citet{glikman2015,glikman2012} are particularly telling: at the peak epoch for galaxy and BH growth, the most luminous quasars are also the most dust reddened, and they are major mergers. This tantalizing discovery points the way forward for the next decade: targeting obscured quasars to get a statistically sound handle on their demographics, measure their SFR and histories to test the co-evolution scenario, not just for the most luminous of quasars but for fainter AGN, and finally study the impact of star-formation and AGN feedback on the host galaxy. The next generation of X-ray, radio, and IR wide field/all sky surveys must be leveraged by efficient (i.e. sensitive /wide aperture/highly multiplexed) spectroscopic surveys in the optical and NIR. Such spectra are needed to obtain, at a minimum, redshifts, and to confirm the MIR/radio/X-ray AGN classifications. To progress in our understanding we need to use absorption and emission lines to constrain the star-formation histories and the physics of the power sources of statistically significant AGN at redshifts $\geq ~1$ with a wide range of obscuration properties. \newline

\begin{figure*}[!h]
\includegraphics[width=0.38\textwidth]{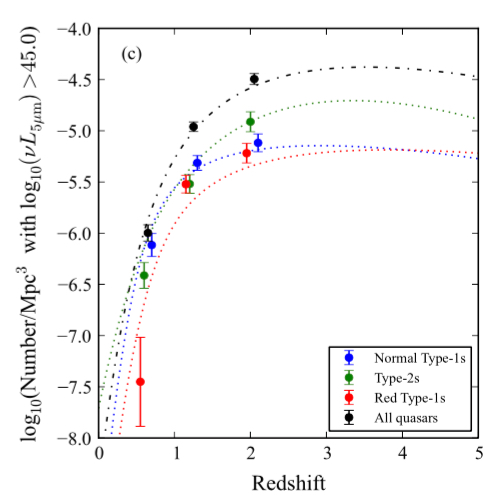}
\includegraphics[width=0.6\textwidth]{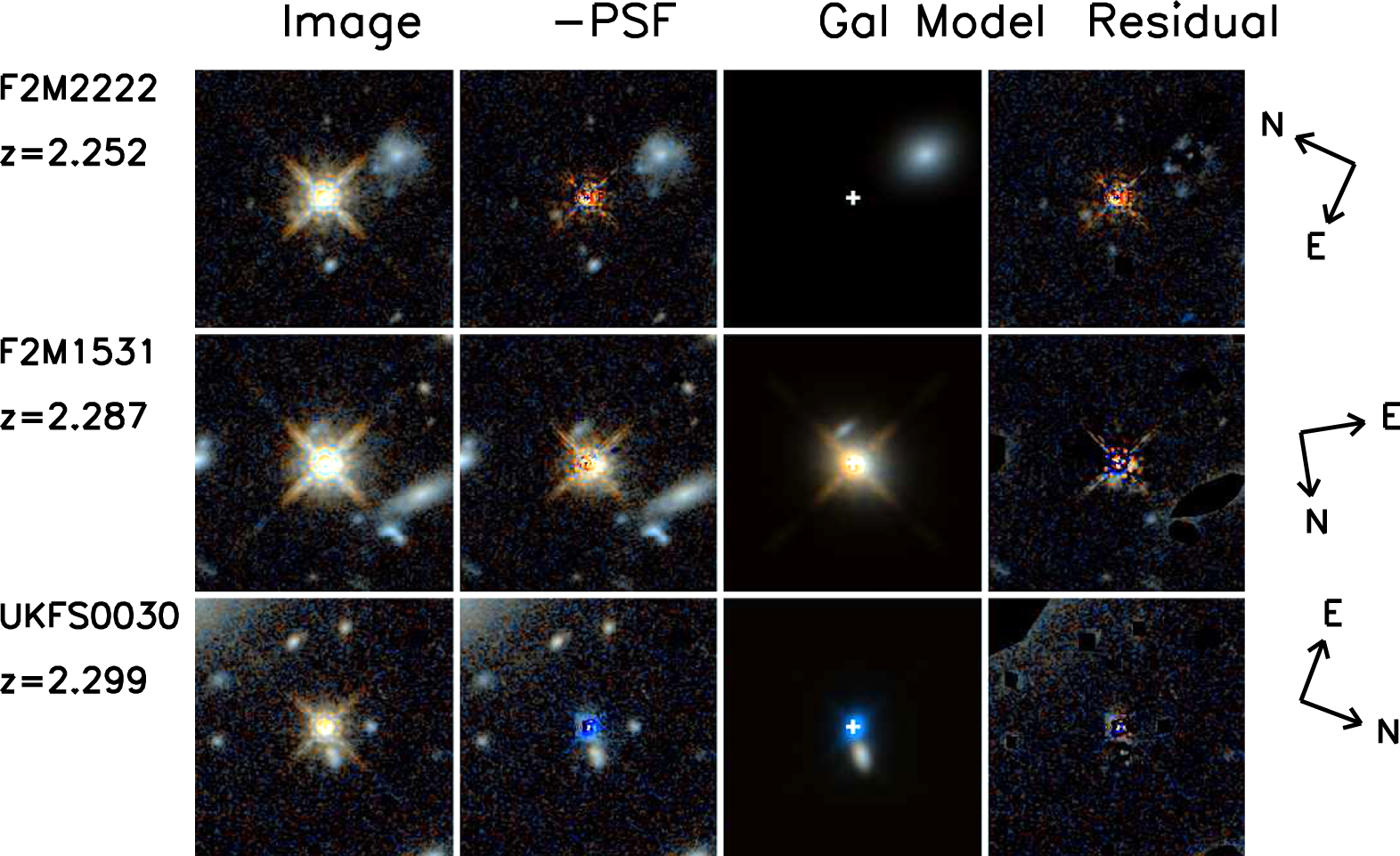}
\caption{({\bf{Left:}}) Figure from \citep{Lacy2015} showing the evolution by type of AGN from a survey of MIR selected AGN. Although only 13 objects of any type at $z\geq 2.8$ were included in this analysis, the difference in evolution between obscured and non-obscured AGN is intriguing and underscores the importance of optical-to-NIR AGN surveys with wide aperture telescopes (for sensitivity), a wide field, and multi-object (for efficient follow up of radio and X-ray deep surveys). ({\bf{Right:}}) Figure 5 from \citep{glikman2015}(reproduced with permission) show HST WFC3/IR images of $z~2$, luminous, red-quasars that suggest the host galaxies are mergers.} 
\label{fig:lacyAGN2015}
\end{figure*}

Demographic studies of AGN populations show that the number density of MIR-selected obscured AGN (with Av $\geq 1$) peaks at a higher redshift (z $\sim 2-3$) than that of unobscured counterparts \citep{Mauduit2012}. \citet{Lacy2015} suggest that there are evolutionary differences between obscured and unobscured sources and speculate that the observed differences between the luminosity functions may be driven by the increased frequency of major mergers of gas rich galaxies at high redshift (see Fig. \ref{fig:lacyAGN2015} ). \citet{Lacy2015} reach those conclusions from optical and NIR spectroscopic observations of MIR selected AGN from wide-field micro-Jansky level surveys with Spitzer IRAC. A spectroscopic facility that can push the sensitivity limit to better probe the high redshift regime z $\sim 3-4$, i.e. large aperture, is therefore desired. \newline

\section{A new Generation of IR, X-ray, and radio-missions to peer through the dust}
The study of obscured AGN in the next decade will be transformed by several IR, X-ray, and radio campaigns which are bound to find millions of AGN. \newline

In the IR the SPHEREx mission \citep{dore2018} will survey the whole sky at a 6" spatial resolution, and a spectral resolution of R$\sim 41-135$ between 0.75 and 5 microns and will be able to isolate AGN based on their MIR colors. However, given its poor angular resolution, SPHEREx will require follow-up observations. A wide-field multi-object spectrometer with optical to NIR coverage, able to detect the faintest and reddest sources in the \citet{Lacy2015} work (i.e. $mAB=23~24$ in 1hr will be ideally suited to follow up SPHEREx's targets. While the SPHEREx mission will perform IR and optical reverberation mapping of the brightest AGN ($i\leq 18$), such studies need to be placed in the context of similar work on fainter AGN. \newline 

The X-ray eROSITA mission \citep{merloni2012} is poised to detect on the order of $10^5$ obscured AGN, thousands of which are expected to be the so far elusive Compton-thick AGN with obscuring column densities greater than $10^{24}~cm^{-2}$ \citep{geor2017,ram2017}. Pinning down the fraction and evolution of those most obscured AGN is essential to our understanding of their evolution. \newline

Finally, several ongoing and planned radio projects will revolutionize our understanding of the role of radio jetted AGN in shaping the environment of both their host galaxy and the circumgalactic medium. The Jansky-VLA/VLASS \citep{lacy2019}, LoFAR/LoTSS\citep{shim2017}, ASKAP/EMU\citep{nor2011}, MeerKAT/MIGHTEE \citep{jar2017} surveys will require wide-field, highly multiplexed spectroscopy to distinguish between high- and low-excitation accretion modes via analysis of emission line ratios. Targeting radio- and MIR-selected AGN with highly multiplexed spectrograph will allow for in-depth studies into the triggering processes for both radiatively efficient and inefficient AGN in the early Universe.

\section{The need for optical to NIR, highly multiplexed spectroscopy }

To obtain a census of accretion we need to measure accretion rates, which require measurements of bolometric luminosities from the AGN. Such estimates require the separation of AGN from SFR emission \citep[e.g.][]{Lani2017}. 
To relate the growth of the central supermassive black hole with that of its host galaxy we also need to estimate the host's star-formation history; this requires spectroscopic follow up in rest-frame UV-optical. \newline

Optical through infrared spectroscopy is also essential for determining the predominant mode of AGN feedback. Two modes of feedback have been invoked to explain why low-redshift massive galaxies are less luminous than cosmological simulations predict: (1) quasar mode feedback at high accretion rates (typical Eddington ratios $>$ 0.01) associated with luminous AGN activity, and (2) radio mode feedback at lower accretion rates, where only low excitation emission lines are typically seen in the rest-frame optical/UV.
In the quasar mode, strong winds from the quasar interact with the ISM of the host galaxy, heating it and preventing star formation. In the radio mode, AGN drive jets and inflate bubbles that heat the circumgalactic and halo gas, which shuts down cooling in massive halos and brings the bright end of the luminosity function into agreement with observations. Radio surveys of AGN \citep[e.g.][]{smol2017} suggest that the kinetic luminosity from radio AGN may be sufficient to balance the radiative cooling of the hot gas at each cosmic epoch since redshift 5, however, these estimates are highly uncertain. \newline

\citet{amar2019} look at a wide range of hydrodynamical simulations and semi-analytic models to estimate the frequency of AGN that are growing via QSO mode, radio mode, and super-Eddington (typically associated with QSOs as well) accretion. The largest differences in the predictions seem to be concentrated in the redshift range 2--4. Most of these models invoke AGN as the primary way of quenching star formation in massive galaxies, and predict similar stellar mass functions and stellar mass densities vs. redshift. This suggests a degeneracy between our understanding of AGN and the effect of massive stars. Detailed studies, (going beyond simple AGN classification) of gas metallicities, ionization conditions, and stellar populations in the hosts of high-$z$ AGN from deep spectroscopy, may be needed to break these degeneracies.\newline


The ages of the stellar populations in the host galaxies are also telling of formation scenarios, and observations of optically selected luminous QSOs seem to show primarily old hosts \citep[e.g.][]{floyd2013}. However, optically selected powerful AGN may not share the same evolutionary path as IR selected powerful AGN \citep{Lacy2015}. \newline


\section{Summary}
X-ray, radio, and IR missions planed for the current and next decade will find millions of AGN, and $\sim 10^5$ obscured AGN. However, to secure AGN classifications, redshifts, star-formation histories and to study triggering and feedback processes for AGN at redshifts when most of the black-hole and galaxy growth happened we need optical to NIR spectroscopic surveys that:
\begin{itemize}
    \item can easily (~6hr) detect the faintest, reddest obscured AGN found in current deepest MIR surveys $\sim$ mAB $\sim$ 24 \citep[i.e.][]{Lacy2015}. This suggests a wide aperture ~10m facility. 
    \item can extend to H-band to reach stellar population diagnostics at $z\geq 1$.
    \item allows for simultaneous observations of thousands of objects to efficiently cover the wide areas mapped by radio and X-ray surveys. 
    \item is a dedicated mission to allow synergies with other AGN and galaxy evolution projects that provide a context in which to understand the nature and evolution of obscured AGN.
\end{itemize}

\pagebreak
\bibliography{WP2020petric}

\begin{thebibliography}{}
\expandafter\ifx\csname natexlab\endcsname\relax\def\natexlab#1{#1}\fi

\bibitem[{{Allevato} {et~al.}(2014){Allevato}, {Finoguenov}, {Civano},
  {Cappelluti}, {Shankar}, {Miyaji}, {Hasinger}, {Gilli}, {Zamorani},
  {Lanzuisi}, {Salvato}, {Elvis}, {Comastri}, \& {Silverman}}]{allevato2014}
{Allevato}, V., {Finoguenov}, A., {Civano}, F., {et~al.} 2014, \apj, 796, 4

\bibitem[{{Amarantidis} {et~al.}(2019){Amarantidis}, {Afonso}, {Messias},
  {Henriques}, {Griffin}, {Lacey}, {Lagos}, {Gonzalez-Perez}, {Dubois},
  {Volonteri}, {Matute}, {Pappalardo}, {Qin}, {Chary}, \& {Norris}}]{amar2019}
{Amarantidis}, S., {Afonso}, J., {Messias}, H., {et~al.} 2019, \mnras,
  arXiv:1902.07982

\bibitem[{{Antonucci}(1993)}]{anton1993}
{Antonucci}, R. 1993, \araa, 31, 473

\bibitem[{{Aretxaga} {et~al.}(1999){Aretxaga}, {Joguet}, {Kunth}, {Melnick}, \&
  {Terlevich}}]{Aretxaga1999}
{Aretxaga}, I., {Joguet}, B., {Kunth}, D., {Melnick}, J., \& {Terlevich}, R.~J.
  1999, \apjl, 519, L123

\bibitem[{{Chiaberge} {et~al.}(2015){Chiaberge}, {Gilli}, {Lotz}, \&
  {Norman}}]{chia2015}
{Chiaberge}, M., {Gilli}, R., {Lotz}, J.~M., \& {Norman}, C. 2015, \apj, 806,
  147

\bibitem[{{Delvecchio} {et~al.}(2014){Delvecchio}, {Gruppioni}, {Pozzi},
  {Berta}, {Zamorani}, {Cimatti}, {Lutz}, {Scott}, {Vignali}, {Cresci},
  {Feltre}, {Cooray}, {Vaccari}, {Fritz}, {Le Floc'h}, {Magnelli}, {Popesso},
  {Oliver}, {Bock}, {Carollo}, {Contini}, {Le F{\'e}vre}, {Lilly}, {Mainieri},
  {Renzini}, \& {Scodeggio}}]{del2014}
{Delvecchio}, I., {Gruppioni}, C., {Pozzi}, F., {et~al.} 2014, \mnras, 439,
  2736

\bibitem[{{Donoso} {et~al.}(2014){Donoso}, {Yan}, {Stern}, \&
  {Assef}}]{donoso2014}
{Donoso}, E., {Yan}, L., {Stern}, D., \& {Assef}, R.~J. 2014, \apj, 789, 44

\bibitem[{{Dor{\'e}} {et~al.}(2018){Dor{\'e}}, {Werner}, {Ashby}, {Bleem},
  {Bock}, {Burt}, {Capak}, {Chang}, {Chaves-Montero}, {Chen}, {Civano},
  {Cleeves}, {Cooray}, {Crill}, {Crossfield}, {Cushing}, {de la Torre},
  {DiMatteo}, {Dvory}, {Dvorkin}, {Espaillat}, {Ferraro}, {Finkbeiner},
  {Greene}, {Hewitt}, {Hogg}, {Huffenberger}, {Jun}, {Ilbert}, {Jeong},
  {Johnson}, {Kim}, {Kirkpatrick}, {Kowalski}, {Korngut}, {Li}, {Lisse},
  {MacGregor}, {Mamajek}, {Mauskopf}, {Melnick}, {M{\'e}nard}, {Neyrinck},
  {{\"O}berg}, {Pisani}, {Rocca}, {Salvato}, {Schaan}, {Scoville}, {Song},
  {Stevens}, {Tenneti}, {Teplitz}, {Tolls}, {Unwin}, {Urry}, {Wandelt},
  {Williams}, {Wilner}, {Windhorst}, {Wolk}, {Yorke}, \& {Zemcov}}]{dore2018}
{Dor{\'e}}, O., {Werner}, M.~W., {Ashby}, M.~L.~N., {et~al.} 2018, arXiv
  e-prints, arXiv:1805.05489

\bibitem[{{Elitzur}(2012)}]{elitzur2012}
{Elitzur}, M. 2012, \apjl, 747, L33

\bibitem[{{Floyd} {et~al.}(2013){Floyd}, {Dunlop}, {Kukula}, {Brown}, {McLure},
  {Baum}, \& {O'Dea}}]{floyd2013}
{Floyd}, D.~J.~E., {Dunlop}, J.~S., {Kukula}, M.~J., {et~al.} 2013, \mnras,
  429, 2

\bibitem[{{Georgakakis} {et~al.}(2017){Georgakakis}, {Aird}, {Schulze},
  {Dwelly}, {Salvato}, {Nandra}, {Merloni}, \& {Schneider}}]{geor2017}
{Georgakakis}, A., {Aird}, J., {Schulze}, A., {et~al.} 2017, \mnras, 471, 1976

\bibitem[{{Glikman} {et~al.}(2015){Glikman}, {Simmons}, {Mailly}, {Schawinski},
  {Urry}, \& {Lacy}}]{glikman2015}
{Glikman}, E., {Simmons}, B., {Mailly}, M., {et~al.} 2015, \apj, 806, 218

\bibitem[{{Glikman} {et~al.}(2012){Glikman}, {Urrutia}, {Lacy}, {Djorgovski},
  {Mahabal}, {Myers}, {Ross}, {Petitjean}, {Ge}, {Schneider}, \&
  {York}}]{glikman2012}
{Glikman}, E., {Urrutia}, T., {Lacy}, M., {et~al.} 2012, \apj, 757, 51

\bibitem[{{Hopkins} {et~al.}(2008){Hopkins}, {Cox}, {Kere{\v s}}, \&
  {Hernquist}}]{hopkins2008}
{Hopkins}, P.~F., {Cox}, T.~J., {Kere{\v s}}, D., \& {Hernquist}, L. 2008,
  \apjs, 175, 390

\bibitem[{{Hopkins} {et~al.}(2007){Hopkins}, {Richards}, \&
  {Hernquist}}]{hopkins2007}
{Hopkins}, P.~F., {Richards}, G.~T., \& {Hernquist}, L. 2007, \apj, 654, 731

\bibitem[{{Jahnke} \& {Macci{\`o}}(2011)}]{jah2011}
{Jahnke}, K., \& {Macci{\`o}}, A.~V. 2011, \apj, 734, 92

\bibitem[{{Jarvis} {et~al.}(2016){Jarvis}, {Taylor}, {Agudo}, {Allison},
  {Deane}, {Frank}, {Gupta}, {Heywood}, {Maddox}, {McAlpine}, {Santos},
  {Scaife}, {Vaccari}, {Zwart}, {Adams}, {Bacon}, {Baker}, {Bassett}, {Best},
  {Beswick}, {Blyth}, {Brown}, {Bruggen}, {Cluver}, {Colafrancesco}, {Cotter},
  {Cress}, {Dav{\'e}}, {Ferrari}, {Hardcastle}, {Hale}, {Harrison}, {Hatfield},
  {Klockner}, {Kolwa}, {Malefahlo}, {Marubini}, {Mauch}, {Moodley}, {Morganti},
  {Norris}, {Peters}, {Prandoni}, {Prescott}, {Oliver}, {Oozeer}, {Rottgering},
  {Seymour}, {Simpson}, {Smirnov}, \& {Smith}}]{jar2017}
{Jarvis}, M., {Taylor}, R., {Agudo}, I., {et~al.} 2016, in Proceedings of
  MeerKAT Science: On the Pathway to the SKA. 25-27 May, 2016 Stellenbosch,
  South Africa (MeerKAT2016). Online at <A
  href=``href=''>href=``https://pos.sissa.it/cgi-bin/reader/conf.cgi?confid=277</A>,
  id.6, 6

\bibitem[{{Juneau} {et~al.}(2011){Juneau}, {Dickinson}, {Alexander}, \&
  {Salim}}]{jun2011}
{Juneau}, S., {Dickinson}, M., {Alexander}, D.~M., \& {Salim}, S. 2011, \apj,
  736, 104

\bibitem[{{Kelly} {et~al.}(2010){Kelly}, {Vestergaard}, {Fan}, {Hopkins},
  {Hernquist}, \& {Siemiginowska}}]{kelly2010}
{Kelly}, B.~C., {Vestergaard}, M., {Fan}, X., {et~al.} 2010, \apj, 719, 1315

\bibitem[{{Kormendy} \& {Ho}(2013)}]{koh2013}
{Kormendy}, J., \& {Ho}, L.~C. 2013, \araa, 51, 511

\bibitem[{{Koutoulidis} {et~al.}(2018){Koutoulidis}, {Georgantopoulos},
  {Mountrichas}, {Plionis}, {Georgakakis}, {Akylas}, \&
  {Rovilos}}]{koutoulidis2018}
{Koutoulidis}, L., {Georgantopoulos}, I., {Mountrichas}, G., {et~al.} 2018,
  \mnras, 481, 3063

\bibitem[{{Lacy}(2019)}]{lacy2019}
{Lacy}, M. 2019, in American Astronomical Society Meeting Abstracts, Vol. 233,
  American Astronomical Society Meeting Abstracts 233, 424.01

\bibitem[{{Lacy} {et~al.}(2015){Lacy}, {Ridgway}, {Sajina}, {Petric}, {Gates},
  {Urrutia}, \& {Storrie-Lombardi}}]{Lacy2015}
{Lacy}, M., {Ridgway}, S.~E., {Sajina}, A., {et~al.} 2015, \apj, 802, 102

\bibitem[{{Lacy} {et~al.}(2018){Lacy}, {Nyland}, {Mao}, {Jagannathan}, {Pforr},
  {Ridgway}, {Afonso}, {Farrah}, {Guarnieri}, {Gonzales-Solares}, {Jarvis},
  {Maraston}, {Nielsen}, {Petric}, {Sajina}, {Surace}, \& {Vaccari}}]{lacy2018}
{Lacy}, M., {Nyland}, K., {Mao}, M., {et~al.} 2018, \apj, 864, 8

\bibitem[{{Lani} {et~al.}(2017){Lani}, {Netzer}, \& {Lutz}}]{Lani2017}
{Lani}, C., {Netzer}, H., \& {Lutz}, D. 2017, \mnras, 471, 59

\bibitem[{{Mart{\'{\i}}nez-Sansigre} \& {Taylor}(2009)}]{mart2009}
{Mart{\'{\i}}nez-Sansigre}, A., \& {Taylor}, A.~M. 2009, \apj, 692, 964

\bibitem[{{Mauduit} {et~al.}(2012){Mauduit}, {Lacy}, {Farrah}, {Surace},
  {Jarvis}, {Oliver}, {Maraston}, {Vaccari}, {Marchetti}, {Zeimann},
  {Gonz{\'a}les-Solares}, {Pforr}, {Petric}, {Henriques}, {Thomas}, {Afonso},
  {Rettura}, {Wilson}, {Falder}, {Geach}, {Huynh}, {Norris}, {Seymour},
  {Richards}, {Stanford}, {Alexander}, {Becker}, {Best}, {Bizzocchi},
  {Bonfield}, {Castro}, {Cava}, {Chapman}, {Christopher}, {Clements}, {Covone},
  {Dubois}, {Dunlop}, {Dyke}, {Edge}, {Ferguson}, {Foucaud}, {Franceschini},
  {Gal}, {Grant}, {Grossi}, {Hatziminaoglou}, {Hickey}, {Hodge}, {Huang},
  {Ivison}, {Kim}, {LeFevre}, {Lehnert}, {Lonsdale}, {Lubin}, {McLure},
  {Messias}, {Mart{\'{\i}}nez-Sansigre}, {Mortier}, {Nielsen}, {Ouchi},
  {Parish}, {Perez-Fournon}, {Pierre}, {Rawlings}, {Readhead}, {Ridgway},
  {Rigopoulou}, {Romer}, {Rosebloom}, {Rottgering}, {Rowan-Robinson}, {Sajina},
  {Simpson}, {Smail}, {Squires}, {Stevens}, {Taylor}, {Trichas}, {Urrutia},
  {van Kampen}, {Verma}, \& {Xu}}]{Mauduit2012}
{Mauduit}, J.-C., {Lacy}, M., {Farrah}, D., {et~al.} 2012, \pasp, 124, 714

\bibitem[{{Merloni} {et~al.}(2012){Merloni}, {Predehl}, {Becker},
  {B{\"o}hringer}, {Boller}, {Brunner}, {Brusa}, {Dennerl}, {Freyberg},
  {Friedrich}, {Georgakakis}, {Haberl}, {Hasinger}, {Meidinger}, {Mohr},
  {Nandra}, {Rau}, {Reiprich}, {Robrade}, {Salvato}, {Santangelo}, {Sasaki},
  {Schwope}, {Wilms}, \& {German eROSITA Consortium}}]{merloni2012}
{Merloni}, A., {Predehl}, P., {Becker}, W., {et~al.} 2012, arXiv e-prints,
  arXiv:1209.3114

\bibitem[{{Miller} {et~al.}(1991){Miller}, {Goodrich}, \& {Mathews}}]{mil91}
{Miller}, J.~S., {Goodrich}, R.~W., \& {Mathews}, W.~G. 1991, \apj, 378, 47

\bibitem[{{Morganti}(2017)}]{morganti2017}
{Morganti}, R. 2017, Frontiers in Astronomy and Space Sciences, 4, 42

\bibitem[{{Morganti} {et~al.}(2013){Morganti}, {Fogasy}, {Paragi}, {Oosterloo},
  \& {Orienti}}]{morganti2013}
{Morganti}, R., {Fogasy}, J., {Paragi}, Z., {Oosterloo}, T., \& {Orienti}, M.
  2013, Science, 341, 1082

\bibitem[{{Norris}(2011)}]{nor2011}
{Norris}, R.~P. 2011, Journal of Astrophysics and Astronomy, 32, 599

\bibitem[{{Ramos Almeida} \& {Ricci}(2017)}]{ram2017}
{Ramos Almeida}, C., \& {Ricci}, C. 2017, Nature Astronomy, 1, 679

\bibitem[{{Shimwell} {et~al.}(2017){Shimwell}, {R{\"o}ttgering}, {Best},
  {Williams}, {Dijkema}, {de Gasperin}, {Hardcastle}, {Heald}, {Hoang},
  {Horneffer}, {Intema}, {Mahony}, {Mandal}, {Mechev}, {Morabito}, {Oonk},
  {Rafferty}, {Retana-Montenegro}, {Sabater}, {Tasse}, {van Weeren},
  {Br{\"u}ggen}, {Brunetti}, {Chy{\.z}y}, {Conway}, {Haverkorn}, {Jackson},
  {Jarvis}, {McKean}, {Miley}, {Morganti}, {White}, {Wise}, {van Bemmel},
  {Beck}, {Brienza}, {Bonafede}, {Calistro Rivera}, {Cassano}, {Clarke},
  {Cseh}, {Deller}, {Drabent}, {van Driel}, {Engels}, {Falcke}, {Ferrari},
  {Fr{\"o}hlich}, {Garrett}, {Harwood}, {Heesen}, {Hoeft}, {Horellou},
  {Israel}, {Kapi{\'n}ska}, {Kunert-Bajraszewska}, {McKay}, {Mohan},
  {Orr{\'u}}, {Pizzo}, {Prandoni}, {Schwarz}, {Shulevski}, {Sipior}, {Smith},
  {Sridhar}, {Steinmetz}, {Stroe}, {Varenius}, {van der Werf}, {Zensus}, \&
  {Zwart}}]{shim2017}
{Shimwell}, T.~W., {R{\"o}ttgering}, H.~J.~A., {Best}, P.~N., {et~al.} 2017,
  \aap, 598, A104

\bibitem[{{Smol{\v c}i{\'c}} {et~al.}(2017){Smol{\v c}i{\'c}}, {Novak},
  {Delvecchio}, {Ceraj}, {Bondi}, {Delhaize}, {Marchesi}, {Murphy},
  {Schinnerer}, {Vardoulaki}, \& {Zamorani}}]{smol2017}
{Smol{\v c}i{\'c}}, V., {Novak}, M., {Delvecchio}, I., {et~al.} 2017, \aap,
  602, A6

\bibitem[{{Soltan}(1982)}]{soltan1982}
{Soltan}, A. 1982, \mnras, 200, 115

\bibitem[{{Urrutia} {et~al.}(2008){Urrutia}, {Lacy}, \& {Becker}}]{urrutia2008}
{Urrutia}, T., {Lacy}, M., \& {Becker}, R.~H. 2008, \apj, 674, 80

\bibitem[{{Zakamska} {et~al.}(2005){Zakamska}, {Schmidt}, {Smith}, {Strauss},
  {Krolik}, {Hall}, {Richards}, {Schneider}, {Brinkmann}, \&
  {Szokoly}}]{zakamska2005}
{Zakamska}, N.~L., {Schmidt}, G.~D., {Smith}, P.~S., {et~al.} 2005, \aj, 129,
  1212

\end{thebibliography}

\end{document}